\documentclass{article}
\usepackage{tikz}
\usepackage{gensymb} 
\usepackage{spconf,amsfonts,amsmath,amsthm,amssymb,graphicx}
\usepackage{adjustbox,lipsum,bm,soul}
\usepackage{cite}
\usepackage{color}
\usepackage{longtable}

\usepackage{algorithm}%
\usepackage{algorithmicx}%
\usepackage{algpseudocode}%

\DeclareMathOperator*{\argmin}{arg\,min}
\newcommand{\Real}{{\mathbb{R}}}
\newcommand{\Complex}{{\mathbb{C}}}

\renewcommand{\vec}[1]{\ensuremath{\boldsymbol{#1}}}
\newcommand{\hvec}[1]{\ensuremath{\Hat{\boldsymbol{#1}}}}

\newcommand{\figref}[1]{Fig.~\ref{fig:#1}}
\newcommand{\tabref}[1]{Table~\ref{tab:#1}}
\newcommand{\eqnref}[1]{Eq.~\ref{eq:#1}}

\usepackage{multirow} 

\title{Deep Image prior with StruCtUred Sparsity (DISCUS) for dynamic MRI reconstruction}
\name{Muhammad Ahmad Sultan, Chong Chen, Yingmin Liu, Xuan Lei, and Rizwan Ahmad\thanks{Corresponding author: Rizwan Ahmad (ahmad.46@osu.edu).}}
\address{The Ohio State University}
\begin{document}
 \abovedisplayskip=3pt 
 \belowdisplayskip=3pt 
 \abovedisplayshortskip=3pt 
 \belowdisplayshortskip=3pt 
 \arraycolsep=3pt

\maketitle
\begin{abstract}
High-quality training data are not always available in dynamic MRI. To address this, we propose a self-supervised deep learning method called deep image prior with structured sparsity (DISCUS) for reconstructing dynamic images. DISCUS is inspired by deep image prior (DIP) and recovers a series of images through joint optimization of network parameters and input code vectors. However, DISCUS additionally encourages group sparsity on frame-specific code vectors to discover the low-dimensional manifold that describes temporal variations across frames. Compared to prior work on manifold learning, DISCUS does not require specifying the manifold dimensionality. We validate DISCUS using three numerical studies. In the first study, we simulate a dynamic Shepp-Logan phantom with frames undergoing random rotations, translations, or both, and demonstrate that DISCUS can discover the dimensionality of the underlying manifold. In the second study, we use data from a realistic late gadolinium enhancement (LGE) phantom to compare DISCUS with compressed sensing (CS) and DIP, and to demonstrate the positive impact of group sparsity. In the third study, we use retrospectively undersampled single-shot LGE data from five patients to compare DISCUS with CS reconstructions. The results from these studies demonstrate that DISCUS outperforms CS and DIP, and that enforcing group sparsity on the code vectors helps discover true manifold dimensionality and provides additional performance gain.


\end{abstract}


\section{Introduction}
\label{sec:intro}
Magnetic resonance imaging (MRI) is a routine diagnostic tool with a broad range of clinical applications. Long scan times, however, are a major hurdle in MRI as they can reduce patient comfort, decrease throughput, and make imaging susceptible to motion artifacts. Consequently, accelerating MRI has been an active area of research \cite{ravishankar2019image}.
More recently, deep learning (DL)-based methods have shown great promise for MRI reconstruction. Several recent studies suggest that such methods can outperform sparsity-based compressed sensing (CS) methods in terms of image quality \cite{zbontar2018fastmri}.

Most DL-based image reconstruction methods require high-quality fully sampled data for supervised training. However, such data are typically not available for dynamic MRI. In contrast, self-supervised methods do not rely on fully sampled data, making them an attractive option for dynamic imaging. Several self-supervised methods have been proposed for MRI reconstruction, including deep image prior (DIP) \cite{ulyanov2018deep}. DIP utilizes the network structure as an implicit prior. For dynamic imaging, Yoo et al. employed DIP by training a generative network to map a low-dimensional manifold to an image series \cite{yoo2021time}. In another study, Hamilton et al. integrated low-rank constraint with DIP to facilitate accelerated free-breathing cardiac imaging \cite{hamilton2023low}. Bell et al. proposed a more robust adaptation of DIP by training the network to function as a denoiser instead of a generator \cite{bell2023robust}. 
 
In this work, we propose an extension of DIP, called deep image prior with structured sparsity (DISCUS). DISCUS trains a single network to map a series of random code vectors to a series of images. In contrast to methods that require specifying the dimensionality of the manifold \cite{yoo2021time}, which may not be known in advance, DISCUS discovers the dimensionality by imposing group sparsity on the code vectors. Also, DISCUS does not assume that temporal closeness (order of acquisition) is tied to image similarity \cite{zou2021dynamic}. This makes DISCUS suitable for free-breathing single-shot applications, where consecutive frames are not necessarily more similar. In the subsequent sections, we describe DISCUS and validate it using simulated and measured data.

\section{Methods}
In this section, we first briefly describe DIP and then use that as a starting point to explain DISCUS.
\subsection{DIP}
\label{DIP}
DIP can be used to solve a wide range of inverse problems \cite{ulyanov2018deep}. To apply DIP to MRI reconstruction, a random code vector $\vec{z}$ is passed through a network $\vec{G}_{\vec{\theta}}$ to generate an estimate of the true image $\vec{x}\in\Complex^N$. 
Both the network parameters, $\vec{\theta}$, and the code vector, $\vec{z}$, are optimized to make the output of the network consistent with the measured k-space data $\vec{y}\in\Complex^M$ through a known forward operator $\vec{A}\in\Complex^{M\times N}$. This optimization process can be expressed as
\vspace{-2mm}

\begin{align} 
\hat{\vec{z}}, \hat{\vec{\theta}} = \argmin_{\vec{z}, \vec{\theta}} {\left\| \vec{A}\vec{G}_{\vec{\theta}}(\vec{z})-\vec{y} \right\|_2^2}.
\label{eq:dip}
\end{align}

Once trained, the final image is recovered by $\hvec{x}=\vec{G}_{\hvec{\theta}}(\hvec{z})$. 


\subsection{DISCUS}
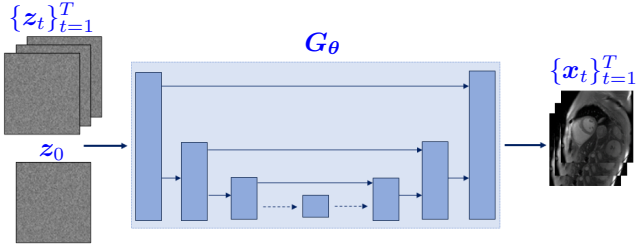
\begin{figure}[h]
  \centering
  \begin{tikzpicture}
    \node[anchor=south west, inner sep=0] (image) at (0,0) {\includegraphics[width=0.49\textwidth]{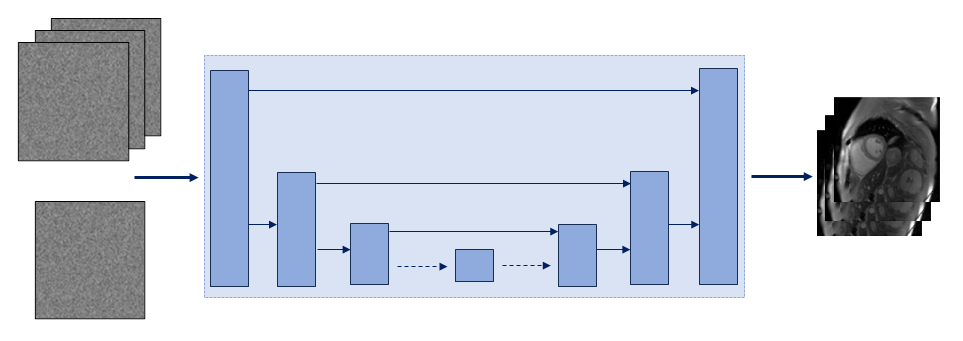}};
    \node[blue,ultra thick] at (0.8,3.2) {$\{\vec{z}_t\}_{t=1}^T$};
    \node[blue,ultra thick] at (0.8,1.4) {$\vec{z}_0$};
    \node[blue,ultra thick] at (4.4,2.85) {$\vec{G}_{\vec{\theta}}$};
    \node[blue,ultra thick] at (8,2.5) {$\{\vec{x}_t\}_{t=1}^T$};
  \end{tikzpicture}
  \caption{Layout of the DISCUS framework. A CNN-based network $\vec{G}_{\vec{\theta}}$ takes one static ($\vec{z}_0$) and one dynamic ($\vec{z}_t$) code vector to generate an estimate of frame $\vec{x}_t$. 
  }
  \label{fig:discus-framewwork}
\end{figure}

The high-level description of DISCUS is provided in \figref{discus-framewwork}. DISCUS attempts to construct an image series with $T$ frames by solving the following optimization problem:
\vspace{-2mm}

{\small
\begin{align}
\hat{\vec{z}}_0, \hat{\vec{z}}_t, \hat{\vec{\theta}} = \argmin_{\vec{z}_0, \vec{z}_t, \vec{\theta}} {\sum_{t=1}^{T}{\left\| \vec{A}_t\vec{G}_{\vec{\theta}}(\vec{z}_0, \vec{z}_t)-\vec{y}_t \right\|_2^2}+\lambda\left\| \vec{z}_{:}\right\|_{2,1}}
\label{eq:discus}
\end{align}}
where $\vec{y}_t\in\Complex^M$, $\vec{x}_t\in\mathbb{C}^N$, and $\vec{A}_t\in\mathbb{C}^{M\times N}$ are measured data, underlying image, and forward operator for the $t^{\text{th}}$ frame, respectively. Also, $\vec{z}_0\in\Real^{kN}$ and $\vec{z}_t\in\Real^{N}$ represent static and dynamic code vectors, respectively, for a user-defined $k>0$. Here,  $\left\| \vec{z}_{:}\right\|_{2,1} = {\sum_{n=1}^N\sqrt{\sum_{t} z_t^2[n]}}$ is a hybrid $\ell_2$-$\ell_1$ norm that first computes elementwise $\ell_2$ norm along the time dimension, followed by the $\ell_1$ norm along the remaining dimensions. Here, $z_t[n]$ represents the $n^{\text{th}}$ element of $\vec{z}_t$.



The static code vector $\vec{z}_0$ is common across all frames, while the dynamic code vector $\vec{z}_t$ is specific to the $t^\text{th}$ image. Code vectors $\vec{z}_0$ and $\vec{z}_t$ are concatenated along the channel dimension and fed into the generator, $\vec{G}_{\vec{\theta}}$. 
Once the network is trained, $\hvec{x}_t = \vec{G}_{\hvec{\theta}} (\hvec{z}_0, \hvec{z}_t)$ produces the final reconstructed image for the $t^{\text{th}}$ frame. 

\subsection{Manifold learning with group sparsity}
 For dynamic MRI, the reconstruction methods based on manifold learning assume that the temporal changes due to respiration or other physiological motions reside on a low-dimensional manifold \cite{poddar2015dynamic}. Manifold learning can be used in the context of DIP \cite{zou2021dynamic}. This is achieved by restricting the dimensionality of the input code vectors to the dimensionality of the manifold and then training the network to map this low-dimensional input to an image series. However, this approach requires that the dimensionality of the manifold is known precisely, which is not always the case for MRI. Changes in image contrast due to RF timing imperfections, heart rate variations, through-plane motion, and bulk motion can alter the dimensionality of the underlying manifold from its nominal value. 

 In DISCUS, we discover the underlying manifold of image series using a hybrid $\ell_2$-$\ell_1$ norm that promotes group sparsity (the second term in \eqnref{discus}) \cite{yuan2006model}. The group sparsity encourages all dynamic code vectors to have a support that is not only sparse but also common across all $T$ code vectors; this support (number of non-zero entries) represents the dimensionality of the underlying manifold. In this work, we do not enforce additional constraints, such as temporal smoothness, on the code vectors \cite{zou2021dynamic} because such constraints do not apply to all dynamic applications, including free-breathing single-shot late gadolinium enhancement (LGE).

\section{Experiments and results}

\subsection{Phantom study}
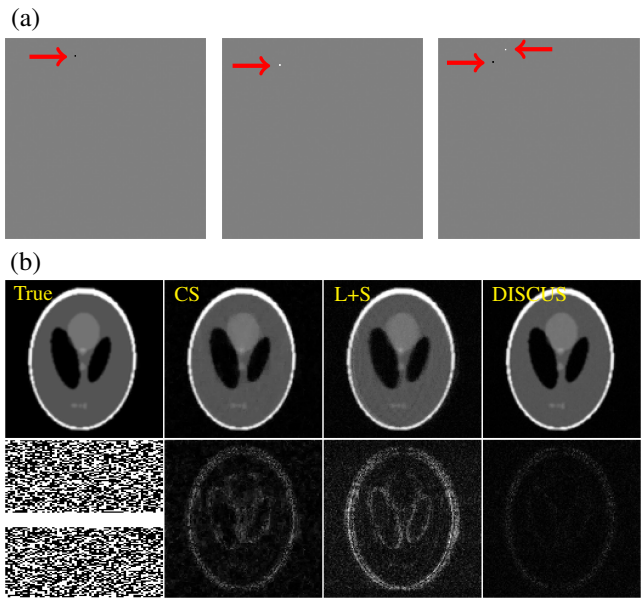
\begin{figure}[!h]
  \centering
  \begin{tikzpicture}

    \node[anchor=south west, inner sep=0] (image) at (0,-0.18) {\includegraphics[width=\columnwidth]{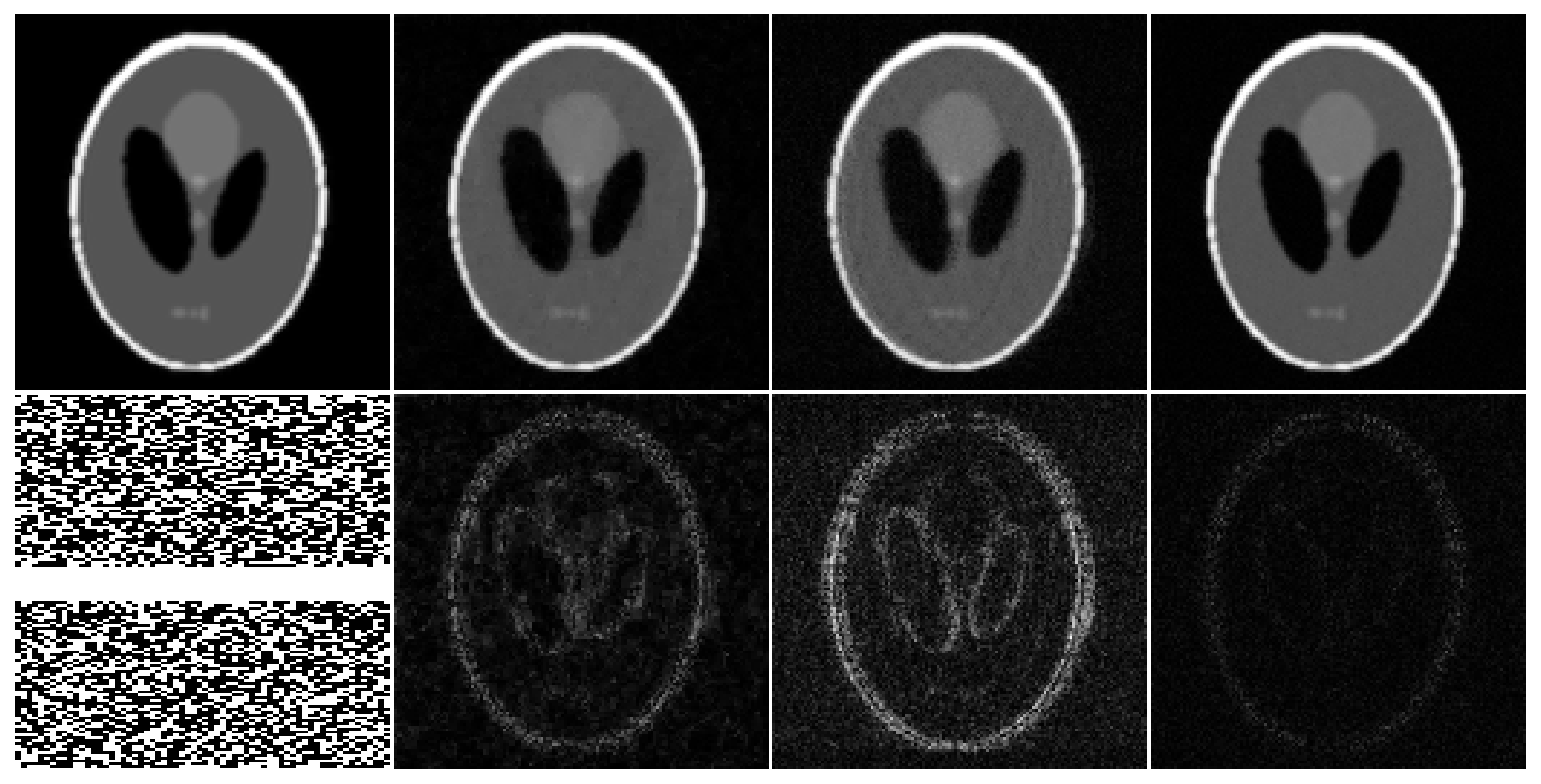}};

    \node[black,ultra thick] at (0.35, 4.35) {(b)};
    \node[yellow, font=\footnotesize] at (0.45, 3.95) {True};
    \node[yellow, font=\footnotesize] at (2.5, 3.95) {CS};
    \node[yellow, font=\footnotesize] at (4.7, 3.95) {L+S};
    \node[yellow, font=\footnotesize] at (7.05, 3.95) {DISCUS};

    \node[anchor=south west, inner sep=0] (image) at (-0.1,4.5) {\includegraphics[width=0.35\columnwidth]{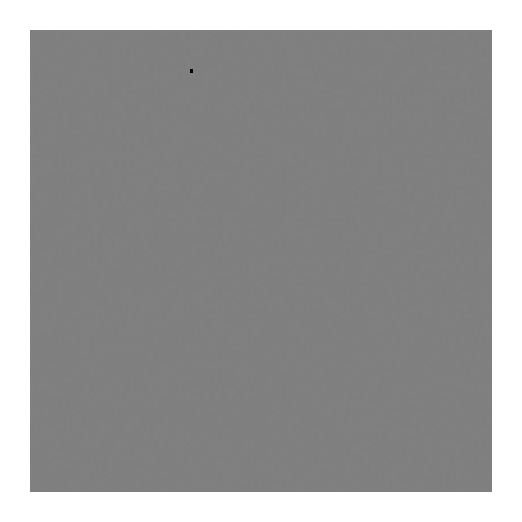}};
    \node[black,ultra thick] at (0.35, 7.6) {(a)};
    \draw[->, ultra thick, red] (0.4,7.1) -- (0.9,7.1);

    \node[anchor=south west, inner sep=0] (image) at (2.78,4.5) {\includegraphics[width=0.35\columnwidth]{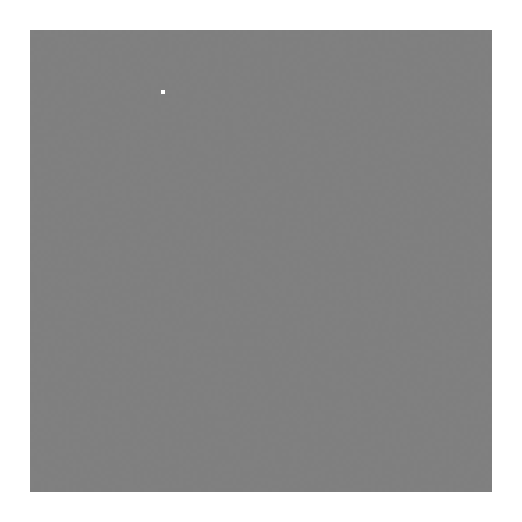}};
    \draw[->, ultra thick, red] (3.1,6.98) -- (3.6,6.98);

    \node[anchor=south west, inner sep=0] (image) at (5.66,4.5) {\includegraphics[width=0.35\columnwidth]{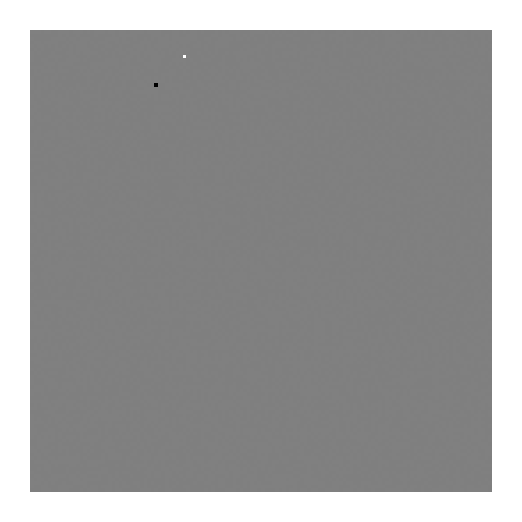}};
    \draw[->, ultra thick, red] (5.95,7.02) -- (6.45,7.02);
    \draw[->, ultra thick, red] (7.35,7.2) -- (6.85,7.2);

  \end{tikzpicture}
  \caption{ (a) One of the dynamic code vectors $\{\hvec{z}_t\}_{t=1}^T$ from image series with rotations (left), translations (middle), and rotations+translations (right). The number of non-zero entries matches the true dimensionality in each image series.  (b) One representative frame from the image series with rotations+translations. The first image in the second row presents the sampling pattern  (frames: left-right,
phase-encoding: top-bottom, frequency-encoding: not shown), and the remaining images show error maps after fivefold amplification.}
  \label{fig:shepp-logan}
\end{figure}

In this study, we demonstrate that DISCUS can discover the low-dimensional manifold. To this end, we simulated a 2D Shepp-Logan phantom of size $128\times 128$. Then, we created three different image series. In the first series, we generated $T=64$ frames, each with a random rotation within  $\pm 3^\circ$ with respect to the first frame. In the second series, we generated $T=64$ frames, each with a random horizontal translation within $\pm3$ pixels with respect to the first frame. In the last series, we generated $T=64$ frames, each with a random rotation and translation. To simulate single-coil k-space data, each frame was Fourier transformed and downsampled using the mask shown in \figref{shepp-logan}b.

Each of the three image series was reconstructed using CS, low-rank + sparse (L+S) \cite{otazo2015low}, and DISCUS. For CS, we imposed $\ell_1$-norm minimization in the spatial wavelet domain. For both CS and L+S, we performed an exhaustive grid search to fine-tune free parameters. For DISCUS, we used a 7-layer U-Net \cite{ulyanov2018deep} and tuned the parameters such as $\lambda$, learning rate, and number of iterations. The tuning parameters across all methods were optimized based on normalized mean squared error (NMSE) defined in dB.



As seen in \figref{shepp-logan}a, DISCUS was able to discover the true manifold dimensionality, defined by the size of the support (non-zero entries) of dynamic code vectors. For the image series with only rotations, only translations, and both rotations and translations, the manifold dimensionality discovered by DISCUS was 1, 1, and 2, respectively. However, we did not observe the locations of the non-zero entries to be of any significance. \tabref{shepp-logan} presents NMSE and structural similarity index (SSIM) values for the three image series, with DISCUS outperforming other methods by a wide margin. 


\begin{table*}[!h] 
\centering
\begin{tabular}{|l|ccc|ccc|ccc|}
\hline
\multirow{2}{*}{\textbf{Metric}} & \multicolumn{3}{c|}{\textbf{Rotation}} & \multicolumn{3}{c|}{\textbf{Translation}} & \multicolumn{3}{c|}{\textbf{Rotation + translation}} \\
& \textbf{CS} & \textbf{L+S} & \textbf{DISCUS} & \textbf{CS} & \textbf{L+S} & \textbf{DISCUS} & \textbf{CS} & \textbf{L+S} & \textbf{DISCUS} \\
\hline
\hline
SSIM & 0.883 & 0.831 & \textbf{0.960} & 0.883 & 0.835 & \textbf{0.962} & 0.882 & 0.834 & \textbf{0.922} \\
\hline
NMSE & -23.82 & -20.30 & \textbf{-31.00} & -23.75 & -20.54 & \textbf{-30.70} & -23.73 & -19.20 & \textbf{-28.67} \\
\hline
\end{tabular}
\caption{NMSE and SSIM values from the Shepp-Logan phantom study.}
\label{tab:shepp-logan}
\end{table*}

\subsection{LGE simulation study}
In this study, we simulated six distinct LGE image series from the MRXCAT phantom \cite{wissmann2014mrxcat}. Each series was cropped to $224 \times 192$ pixels and had $T=32$ frames with $8$ receive coils. All data were retrospectively downsampled at $R=4$ using a realistic golden ratio offset (GRO) Cartesian sampling mask with fully sampled readout \cite{joshi2022technical}. The parameters defining tissue contrast, including proton density, T1, inversion time, and gadolinium concentration, were selected to match routine LGE scans at our institute. To incorporate breathing-induced motion, a unique breathing pattern was simulated for each image series. To make the study more realistic, a myocardial scar was simulated in three of the six image series. White Gaussian noise was added to the k-space data to yield a signal-to-noise ratio (SNR) of 25 dB. The coil-combined reference from the fully sampled data was used for the quantitative assessment. The network in DISCUS was trained for 12,000 iterations, requiring 90 minutes on a single NVIDIA RTX3090 GPU.

\begin{table}[!h]
\smallskip
\centering
\begin{adjustbox}{width=0.47\textwidth}
\begin{tabular}{|l|l|l|l|l|l|}
\hline
 & \textbf{CS} & \textbf{L+S} & \textbf{DIP} & \textbf{DISCUS-GS} & \textbf{DISCUS}\\ \hline \hline
SSIM  & 0.905    & 0.922 & 0.846 & 0.958 & \bf{0.978} \\ \hline
NMSE  & -24.25    & -22.44 & -21.66 & -27.06  & \bf{-28.03} \\ \hline
\end{tabular}
\end{adjustbox}
\caption{NMSE and SSIM values for the simulated LGE data. The numbers are averaged over six distinct image series.}
\label{tab:MRXCAT-comparison}
\end{table}

\begin{figure*}[!h]
  \centering
  \begin{tikzpicture}
    \node[anchor=north west, inner sep=0] (image) at (0,0) {\includegraphics[width=1\textwidth]{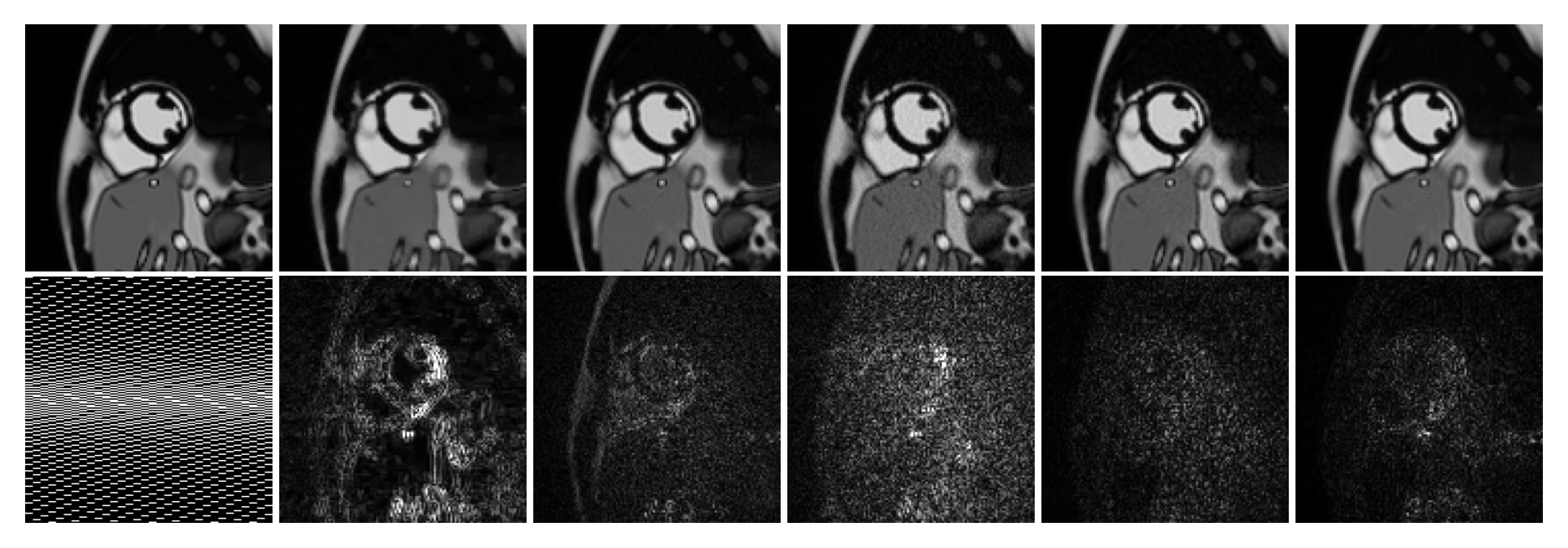}};

    \node[yellow, fill=black] at (0.9,-0.55) {True};
    \node[yellow, fill=black] at (3.6,-0.55) {CS};
    \node[yellow, fill=black] at (6.6,-0.55) {L+S};
    \node[yellow, fill=black] at (9.5,-0.55) {DIP};
    \node[yellow, fill=black] at (13.1,-0.55) {DISCUS-GS};
    \node[yellow, fill=black] at (15.65,-0.55) {DISCUS};
  \end{tikzpicture}
  \caption{Top row: A representative frame from a simulated LGE image series with a pronounced myocardial scar. Bottom row: sampling mask (frames: left-right, phase-encoding: top-bottom, frequency-encoding: not shown) and $\times 8$ error maps.}
  \label{fig:MRXCAT}
\end{figure*}
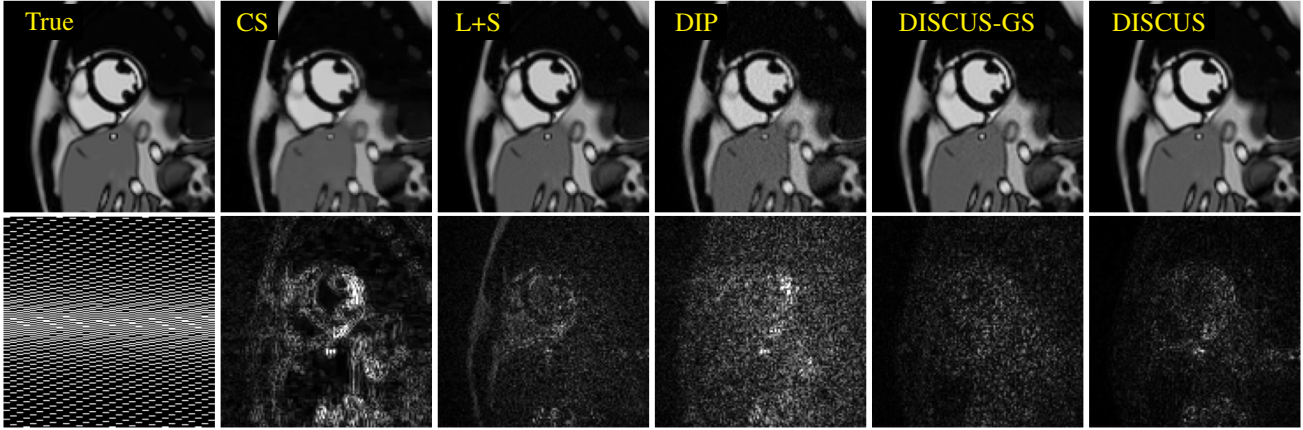

We performed reconstructions using CS, L+S, DIP, DISCUS, and DISCUS without group sparsity (DISCUS-GS). These methods were compared in terms of NMSE and SSIM, as summarized in \tabref{MRXCAT-comparison}. DISCUS outperforms CS, L+S, and DIP by a wide margin for both NMSE and SSIM. These results also show that removing group sparsity worsens the performance by approximately one dB. \figref{MRXCAT} shows a representative frame from one of the image series. Compared to DISCUS, CS exhibits excessive blurring around the myocardial scar, DIP exhibits excessive noise, and L+S exhibits a more structured error map.

\subsection{LGE patient study}
Five free-breathing image series, each with $T=32$ frames, were collected from patients on a 1.5T scanner (MAGNETOM Sola, Siemens Healthcare, Erlangen, Germany) using phase-sensitive inversion recovery LGE sequence \cite{kellman2002phase}. To enable acquisition without acceleration, i.e., $R=1$, the data were collected with a temporal footprint of 249.6 ms and a compromised in-plane spatial resolution of $2.4 \times 2.9$ mm$^2$. The data were compressed to eight virtual coils. The coil sensitivities were estimated using ESPIRiT \cite{shin2014calibrationless}, followed by intensity correction \cite{lei2023surface}. Each image series was retrospectively undersampled using the GRO sampling mask at $R=2,3,$ and $4$ \cite{joshi2022technical}. The coil-combined reference from the fully sampled data was used for the quantitative assessment. The network in DISCUS was trained for 12,000 iterations, requiring 75 minutes on a single NVIDIA RTX3090 GPU.

\begin{table*}[!h]
    \centering
    \begin{tabular}{|l|ccc|ccc|ccc|}
        \hline
        \multirow{2}{*}[0pt]{\textbf{Metric}} & \multicolumn{3}{c|}{\textbf{R=2}} & \multicolumn{3}{c|}{\textbf{R=3}} & \multicolumn{3}{c|}{\textbf{R=4}} \\ 
        & \textbf{CS} & \textbf{L+S} & \textbf{DISCUS} & \textbf{CS} & \textbf{L+S} & \textbf{DISCUS} & \textbf{CS} & \textbf{L+S} & \textbf{DISCUS} \\
        \hline
        \hline
        SSIM & 0.977 & \textbf{0.984} & 0.982 & 0.958 & \textbf{0.983} & 0.980 & 0.933 & \textbf{0.980} & 0.978 \\
        \hline
        NMSE & -23.31 & -20.21 & \textbf{-25.03} & -20.00 & -19.49 & \textbf{-23.53} & -17.78 & -19.08 & \textbf{-22.86} \\
        \hline
    \end{tabular}
    \caption{NMSE and SSIM values for the patient LGE data. The numbers are averaged over five image series.}
    \label{tab:lge}
\end{table*}

\begin{figure*}[!h]
  \centering
  \begin{tikzpicture}
    \node[anchor=north west, inner sep=0] (image) at (0,0) {
    \resizebox{1\textwidth}{1.2\height}{\includegraphics{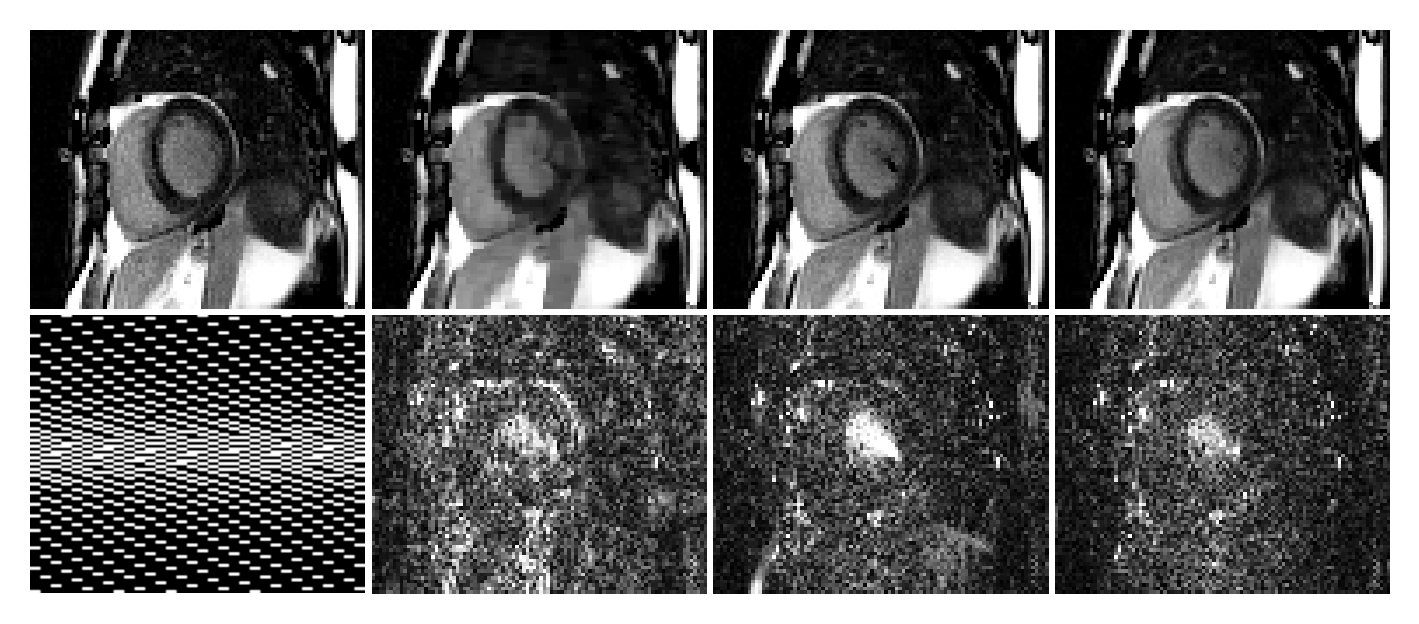}}
};

    \node[yellow, fill=black] at (0.9,-0.55) {True};
    \node[yellow, fill=black] at (5.1,-0.55) {CS};
    \node[yellow, fill=black] at (9.5,-0.55) {L+S};
    \node[yellow, fill=black] at (14.2,-0.55) {DISCUS};
    \draw[->, very thick, red] (5.605,-1.6) -- (6.05,-1.6);
    \draw[->, very thick, red] (10.55,-1.7) -- (10.95,-1.7);
  \end{tikzpicture}
  \caption{Top row: A representative frame from one of the patients at $R=4$. Bottom row: sampling mask (frames: left-right, phase-encoding: top-bottom, frequency-encoding: not shown) and $\times 5$ error maps.}
  \label{fig:lge}
\end{figure*}
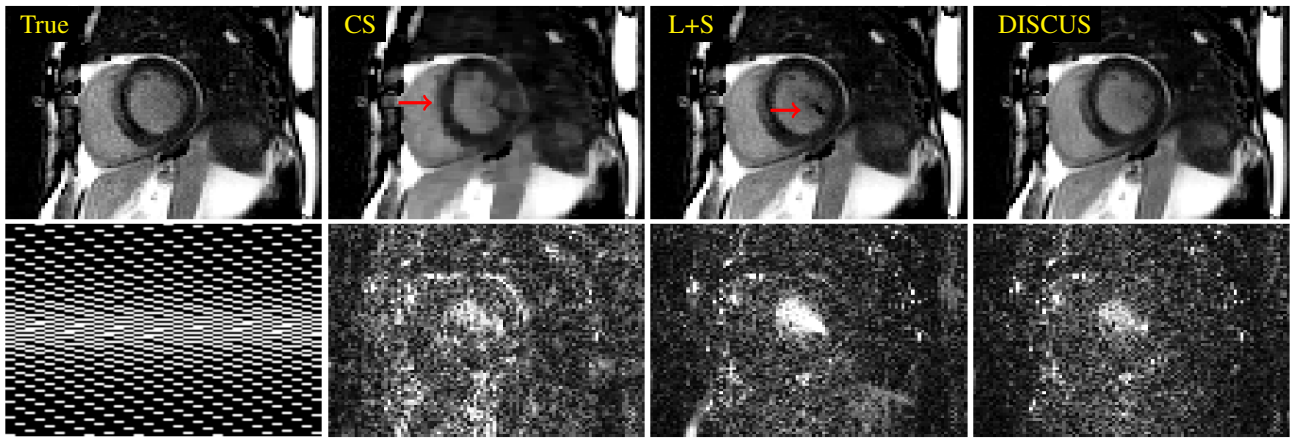

\tabref{lge} compares DISCUS with CS and L+S at three different acceleration rates. In terms of NMSE, DISCUS offers a significant advantage over CS and L+S at all three acceleration rates, while L+S marginally outperforms DISCUS in terms of SSIM. \figref{lge} shows a representative frame from one of the image series at $R=4$. DISCUS is able to preserve fine details while CS shows excessive smoothing and blocky artifacts and L+S shows a significant artifact inside the left ventricle, as highlighted by red arrows.

\section{Conclusions}
We have proposed a self-supervised reconstruction method, DISCUS, for dynamic MRI. DISCUS learns the underlying manifold by promoting group sparsity among frame-specific code vectors. This feature enables DISCUS to discover the low-dimensional manifold even when the manifold dimensionality is not known in advance. Using simulated and measured MRI data for free-breathing single-shot LGE, we demonstrate that DISCUS outperforms CS and DIP in terms of image quality. Our future efforts will focus on applying DISCUS to prospectively undersampled data and to applications other than LGE. 
\clearpage

\section{Additional Information}
\textbf{Acknowledgment}: This work was funded by NIH grants R01HL135489, R01HL151697, and R01EB029957.
\vspace{2mm}

\noindent
\textbf{Compliance with ethical standards}: This study was performed in line with the principles of the Declaration of Helsinki. Approval was granted by the Institutional Review Board (IRB) of The Ohio State University (2019H0076).




\bibliographystyle{IEEEbib}
\bibliography{root.bib}

\end{document}